\documentclass{article}

\usepackage[T1]{fontenc}
\usepackage{array}
\usepackage{graphics}
\usepackage{amssymb,amsfonts,amsmath,stmaryrd}
\usepackage{cite,enumerate,float,indentfirst}
\usepackage{color}
\usepackage{ytableau}
\usepackage{tikz}
\usetikzlibrary{arrows,snakes,backgrounds}

\def\be{\begin{eqnarray}}
\def\ee{\end{eqnarray}}
\def\nn{\nonumber}

\def\p{\partial}
\def\tr{{\rm tr}\,}

\definecolor{red}{rgb}{1,0,0}
\definecolor{orange}{rgb}{1,0.5,0}
\definecolor{violet}{rgb}{0.7,0,1}

\hoffset= - 1.0in         

\begin{document}

\begin{center}
\begin{small}
\hfill MIPT/TH-12/22\\
\hfill FIAN/TD-05/22\\
\hfill ITEP/TH-15/22\\
\hfill IITP/TH-13/22\\

\end{small}
\end{center}

\vspace{.5cm}

\begin{center}
\begin{Large}\fontfamily{cmss}
\fontsize{15pt}{27pt}
\selectfont
	\textbf{Bilinear character correlators in superintegrable theory}
	\end{Large}
	
\bigskip \bigskip

\begin{large}
A. Mironov$^{a,b,c,}$\footnote{mironov@lpi.ru; mironov@itep.ru},
A. Morozov$^{d,b,c,}$\footnote{morozov@itep.ru},
\end{large}

\bigskip

\begin{small}
$^a$ {\it Lebedev Physics Institute, Moscow 119991, Russia}\\
$^b$ {\it Institute for Information Transmission Problems, Moscow 127994, Russia}\\
$^c$ {\it NRC ``Kurchatov Institute'' - ITEP, Moscow 117218, Russia}\\
$^d$ {\it MIPT, Dolgoprudny, 141701, Russia}
\end{small}
 \end{center}

\bigskip

\begin{abstract}
We continue investigating the superintegrability property of matrix models,
i.e. factorization of the matrix model averages of characters.
This paper focuses on the Gaussian Hermitian example, where the role of characters
is played by the Schur functions.
We find a new intriguing corollary of superintegrability:
factorization of an infinite set of correlators {\it bilinear} in the Schur functions.
More exactly, these are correlators of products of the Schur functions and polynomials $K_\Delta$ 
that form a complete basis in the space of invariant matrix polynomials. 
Factorization of these correlators with a small subset of these $K_\Delta$ follow from the fact that the Schur functions
are eigenfunctions of the generalized cut-an-join operators,
but the full set of $K_\Delta$ is generated by another infinite commutative set of operators, which we manifestly describe.
\end{abstract}

\bigskip

\section{Introduction}

In celestial mechanics,
superintegrability (SI) implies existence of an additional operator (Laplace-Runge-Lenz vector)
which commutes with the Hamiltonian
and is somehow different (superficial) as compared to the ``obvious'' commuting set of operators
(rotations), which are responsible for the ordinary integrability.

In the case of matrix models, even this language is still to be developed:
our original definition of superintegrability in \cite{IMM,MM} (based on the
phenomenon earlier observed in \cite{DiF}--\cite{Pop}, see also some preliminary results in \cite{Kaz}--\cite{MKR} and later progress in \cite{MMten}--\cite{MO},\cite{MMrev,MMd}) implies the mapping
between a big space ${\cal X}$ (functions of matrix eigenvalues or time-variables $p_k$)
to a small one, ${\cal Z}$ (functions of the matrix {\it size} $N$)
with the Schur functions $S_R$ being a kind of eigenfunctions of this
contraction map.
Despite the setting looks very poor, the phenomenon clearly exists:
a minor deformation of the Gaussian measure (which preserves integrability)
is not compensated by a small
deformation of the Schur functions so that the {\it super}integrability
property $\Big<S_R\{p\}\Big> \ \sim\ S_R\{N\}$ is preserved.
Obviously, the setting should be lifted to the case when the both spaces,
${\cal X}$ and ${\cal Z}$ are ``equal''.
This could mean, for instance, that one can consider the Schur functions
not just as a subjects of averaging over matrices, but rather as common eigenfunctions
of a commuting set of operators.
In fact, integrability and superintegrability are both related to existence
of different mutually commuting operators.
But how to distinguish between different operators,
and separate them into sets
which are responsible respectively for integrability
and for the {\it super}integrability?

A related question can be what is the reason for an additional enhancement in the case
of Dotsenko-Fateev (double logarithmic) measure,
where one gets a whole set of factorized Kadell integrals
with not only averages of Schur functions,
but also of their peculiar multilinear combinations being described by nice factorization formulas
(giving rise to Nekrasov functions) \cite{Kadell,MMrev,MMd}?

In this letter, we argue that these two questions are related to each other, and demonstrate that, in the Gaussian Hermitian
model case, factorization of bilinear averages of Schur functions, which encodes the superintegrability is also due to existence
of an infinite set of commuting operators. More exactly, we demonstrate that there is a complete set of symmetric polynomials
$K_\Delta$ such that the averages $\Big<K_\Delta S_R\Big>$ are proportional to the averages $\Big<S_R\Big>$:
\be\label{K}
\Big<K_\Delta S_R\Big>=\mu_{\Delta,R}\Big<S_R\Big>
\ee 
where the eigenvalues $\mu_{\Delta,R}$ do not depend on $N$. This set of polynomials $K_\Delta$ is generated by an infinite set of 
commuting operators $W^-_\Delta$.

The letter is organized as follows. In sections 2 and 3, we discuss the generic property of superintegrability. In section 4, we consider the set of cut-and-join operators $W_\Delta$ \cite{MMN} as a natural candidate for the infinite set of commuting operators generating $K_\Delta$, and realize that it gives rise to only part of $K_\Delta$. Hence, in section 5, we construct the set $W^-_\Delta$ that generates all $K_\Delta$. In section 6, we discuss examples of eigenvalues $\mu_{\Delta,R}$, and, in section 7, we find an explicit formula for $\mu_{\Delta,R}$.
Section 8 contains concluding remarks, and, in the Appendix, we explicitly list polynomials $K_\Delta$ for all $\Delta$ up to level 6.

\paragraph{Notation.} The Schur functions are symmetric polynomials of variables $x_i$, $i=1,\ldots,N$. In particular, $x_i$ can be eigenvalues of a matrix $H$. We denote through $S_R\{p_k\}$ the Schur functions as functions of power sums $p_k=\sum_ix_i^k$. 
When we emphasize that $x_i$ are the eigenvalues of $H$, we use the notation $P_k:=\tr H^k$.

The Schur function depends on the partition (Young diagram) $R$, which is a set of lines with lengths $R_1\ge R_2\ge\ldots\ge R_{l_R}$.
We also denote through $S_{R/T}$ the skew Schur functions, and sometimes use the notation
\be
d_R:=S_R\{\delta_{k,1}\}
\ee

\section{SI in matrix models}

According to \cite{MM}, see also \cite{MMrev} and references therein,
SI means that there exits a linear basis in the space of observables
such that all the elements of the basis have ``very simple'' averages.
In practice, this ``very simple'' means fully factorized.
Moreover, this distinguished basis is usually formed by characters
of an underlying symmetry algebra (to which the matrices belong),
and the average of each character is again just the same character, only at
a different (diminished) space of variables.
The typical example is the Gaussian Hermitian model, where averages over
Hermitian matrices are defined
\be
\left<F(H)\right> := \int F(H) e^{-\frac{1}{2}\tr H^2 } dH
\ee
$dH$ being the Haar measure on Hermitian matrices normalized in such a way that $\Big<1\Big>$.

If the function $F$ is invariant, i.e. depends on the eigenvalues $h_i$ of $H$, one can integrate over angular variables, and
\be
\left<F(H)\right> =
\int_{-\infty}^\infty \prod_{i<j}^N F(h_i)(h_i-h_j)^2 \prod_{i=1}^N e^{-\frac{1}{2}h_i^2 } dh_i
\ee
SI in this case states that averages of the Schur functions $S_R\{P_k=\tr H^k\}$
are
\be
\boxed{
\Big< S_R \Big> \ =\  \frac{S_R\{N\}\cdot S_R\{\delta_{k,2}\}}{S_R\{\delta_{k,1}\}}
}
\label{siherm}
\ee
At the r.h.s. are the same Schur functions but at very special points:
the main one is $S_R\{N\}:=S_R\{p_k=N\}$.

There are very similar statements $<character>\ \sim character$ for a variety of other
eigenvalue models, see \cite{MMrev} for an extensive list.

\section{Does SI really exist in matrix models?}

A natural question is if there is any true sense in the above observation?
Perhaps, one can always find such a distinguished basis?
It is therefore instructive to look at what happens in the same Hermitian model
when one changes the background potential from the Gaussian one to anything else.

The Gaussian partition function
\be
Z\{p\} = \Big< e^{\sum_k \frac{p_kP_k}{k}}\Big>
= \sum_R S_R\{p\} \Big<S_R\{P\}\Big>
\label{Gpf}
\ee
allows one to define an average in arbitrary non-Gaussian potential $\sum_k T_k X^k$:
\be
Z^{(T)}\{p\} = \Big< e^{\sum_k \frac{(p_k+T_k)P_k}{k}}\Big>
= \sum_Q S_Q\{p+T\} \Big<S_Q\{P\}\Big>
= \sum_R S_R\{p\} \sum_Q S_{Q/R}\{T\}\Big<S_Q\{P\}\Big>
\ee
On the other hand, one can rewrite it as a sum over averages in the $T$-background,
\be
\Big< e^{\sum_k \frac{(p_k+T_k)P_k}{k}}\Big>=\sum_RS_R\{p\}\Big<S_R\{P\}\Big>^{(T)}
\ee
i.e. the $T$-deformed averages (note that the normalization is still Gaussian, i.e. the definition of average is not changed) are
\be\label{Tb}
\boxed{
\Big<S_R\{P\}\Big>^{(T)} =
\sum_Q S_{Q/R}\{T\}\Big<S_Q\{P\}\Big>
}
\ee
These averages are non-factorized infinite series that, actually, can not to be simplified
even for finitely many non-vanishing $T_k$.
They much more complicated as compared to
$ \Big<S_R\{P\}\Big> = \frac{S_R[N]\cdot S_R\{\delta_{k,2}\}}{S_R\{\delta_{k,1}\}}$,
and there is no any (obvious) way to modify the Schur functions in order to produce factorized averages,
not to say that the preferred basis, even if existed,
would not be formed by the characters of $sl_N$.
The only exceptions are the deformation by $T_1$ and $T_2$ only,
which preserve Gaussianity.

In this sense, what we call  {\it super}integrability is an obviously non-trivial feature,
which, in this concrete example, distinguishes the Gaussian potential among the arbitrary ones.
Note that it is clearly a further restriction as compared to the ordinary integrability,
the latter one is preserved by arbitrary $T$-deformations
and does {\it not} require Gaussianity:
all $Z^{(T)}\{p\}$ are KP $\tau$-functions,
just they are associated with $T$-dependent points of the universal Grassmannian.
Thus, {\bf superintegrability exists and is a strong refinement of ordinary integrability}.

\section{Constructing $K_\Delta$: $W$-operators}

In the next sections, we assume that
Schur functions are restricted to the Miwa locus
$S_R\{p_k=\tr H^k\}$ with $N\times N$ matrix $H$.
It is a little less general than arbitrary time variables,
but still far away from restricting the Schur functions to their
Gaussian averages $\Big<S_R\Big>$.
We will assume that $N\geq |R|$ though this restriction is not necessary, and
formulas are basically correct at any $N$: one just has to be careful with normalizations.
For instance, $\mu_{\Delta,R}$ is a ratio of two zeroes unless $N\geq |R|$.
In a proper normalization, both sides of formulas typically vanish unless $N\geq |R|$.

Since the Schur functions are common eigenfunctions \cite{MMN} of  the operators
\be
\hat W_\Delta := \ :\prod_{a=1}^{l_\Delta}  \tr \left(H\frac{\p}{\p H_{tr}}\right)^{\Delta_a}:
\ee
\be
\hat W_\Delta S_R = \lambda_{\Delta,R} S_R
\label{SefW}
\ee
where the eigenvalues are appropriately normalized symmetric-group characters,
$\lambda_{\Delta,R}=\varphi_R(\Delta)$ \cite{MMN}, and the normal ordering $:\ldots :$
implies all the derivatives put to the right.
One can use integration by parts to get
\be
\left< S_R \cdot \left(
e^{\frac{1}{2}\tr H^2}\, \hat W_\Delta^\dagger\,
e^{-\frac{1}{2}\tr H^2}\right)\right>
= \lambda_{\Delta,R}\cdot \Big<S_R\Big>
\label{WonS}
\ee
Since the expression in brackets at the l.h.s. is a polynomial in $H$, it can be expanded into a linear combination of the Schur functions,
\be
\boxed{
\Big<\left(\sum_{|Q|\leq 2|\Delta|}  C_{\Delta Q}(N)\ S_Q\right)\cdot S_R\Big>
= \lambda_{\Delta, R}\cdot\Big<  S_R\Big>
}
\label{parcorG}
\ee
with
\be
\sum_Q C_{\Delta Q}(N)\cdot S_Q\{p_k=\tr H^k\}
= e^{\frac{1}{2}\tr H^2}\, \hat W_\Delta^\dagger\, e^{-\frac{1}{2}\tr H^2}
= e^{\frac{1}{2}\tr H^2}\ddag \prod_{a=1}^{l_\Delta}  \tr \left(-\frac{\p}{\p H_{tr}}H\right)^{\Delta_a}
\ddag
e^{-\frac{1}{2}\tr H^2}
\ee
where the normal ordering $\ddag\ldots\ddag$ this time implies that all the derivatives are put to the left.

In particular, for $\hat W_{[1]}$, which is just a dilatation operator with $\lambda_{[1],R} = |R|$,
this means:
\be\label{K2}
\Big< (P_2-N^2)\cdot S_R\{P\} \Big> =|R|\cdot \Big<S_R\{P\}\Big>
\ee
which is indeed true
(for example, one can use that $p_2S_{[2r]} = S_{[2r+2]} + S_{[2r,2]}-S_{[2r,1,1]}$
and similar relations for non-symmetric representations $R$). For example,
\be
\hat W_{[2]}: &  \Big<(P_4-4NP_2-P_1^2+2N^3)\cdot S_R\Big> = \lambda_{[2],R}\cdot \Big< S_R\Big>
\nn \\
\hat W_{[1,1]}: &
\Big<\left(P_2^2-(2N^2+3)P_2 +N^2(N^2+1)\right)\cdot S_R\Big>
= \lambda_{[1,1],R}\cdot \Big< S_R\Big>
\nn \\
\hat W_{[3]}: &  \Big<\left(-P_6 + 6NP_4 + 3P_3P_1  + 3P_2^2
- (15N^2 + 6)P_2 - 6NP_1^2 + 5N^4 + N^2\right)\cdot S_R\Big> = \lambda_{[3],R}\cdot \Big< S_R\Big>
\nn \\
\ldots
\label{WonS1}
\ee

Equations (\ref{parcorG}) are rather poor --
they are {\it not} sufficient to express all Gaussian pair correlators,
they are just sum rules, which impose certain constraints on them.
This is because the number of Young diagrams $\#_{2n} > \#_n$,
the former number is what we need for complete set of pair correlators,
the latter number is what we can actually deduce from (\ref{SefW}).

\section{Constructing $K_\Delta$: $W^-$-operators}

In this section, we discuss that, in order to construct the full set of operators, one has to consider another set of commuting operators, which are a kind of ``lowering'' operators in the space of Schur functions.

Let us note that, in addition to relations (\ref{WonS1}), there more bilinear Schur averages of the (\ref{K}) type: for instance, there is the relation
\be\label{K11}
\Big< (P_1^2-N)\cdot S_R \Big> =
\Big< (S_{[2]}+S_{[1,1]}-N )\cdot S_R \Big> =
\mu_{[1,1],R}\cdot \Big<S_R\Big>
\ee
The l.h.s. of this formula would appear if we act on
$e^{-\frac{1}{2}\tr H^2}$ with the operator $(\tr \frac{\p}{\p H})^2$.
This operator with
$\tr \frac{\p}{\p H} = N\frac{\p}{\p p_1} + \sum_{k=2}^\infty p_{k-1}\frac{\p}{\p p_k}$
does not have $S_R$ as an eigenfunction,
\be
\left(\tr \frac{\p}{\p H}\right)^2 S_R \neq \mu_{[1,1],R} \cdot  S_R
\label{noef}
\ee
What happens is that its action is conspired with the SI formula:
despite (\ref{noef}) forbids $S_R$ to be an eigenfunction,
i.e. equation does not hold at the ``operator level'',
it does hold for the Gaussian averages:
\be
\Big<\left(\tr \frac{\p}{\p H}\right)^2 S_R\Big> \ =  \mu_{[1,1],R} \cdot  \Big<S_R\Big>
\label{yesavef}
\ee
Only for restricted set (\ref{WonS1}) they are promoted to the operator level (\ref{SefW}).

Now our main claim is that one can construct in a similar way {\bf the full set of polynomials $K_\Delta$ for (\ref{K})}. That is, define
\be\label{Wd}
\hat W^-_k:=\tr\left( \frac{\p}{\p H}\right)^k\nn\\
\hat W^-_\Delta:=\prod_a^{l_\Delta}W^-_{\Delta_a}
\ee
Then,
\be\label{W}
\boxed{
\Big<\hat W^-_\Delta\ S_R\Big>=\Big<K_\Delta\cdot S_R\Big>=\mu_{\Delta,R}\cdot  \Big<S_R\Big>
}
\ee
where
\be\label{KD}
\boxed{
\phantom{\Big<}K_\Delta=e^{\frac{1}{2}\tr H^2}\, \hat W^-_\Delta\, e^{-\frac{1}{2}\tr H^2}
}
\ee
Note that
\be
K_\Delta=P_\Delta+\hbox{lower degrees}
\ee
so that they form a complete set polynomials at any given level.

These formulas can be considered as {\bf  one more reformulation (avatar) of superintegrability}. As we already discussed in the Introduction, it is related to an infinite set of commuting operators $W^-_\Delta$, which are manifestly given by (\ref{Wd}).

In section 7,  we prove these relations, and find explicit expressions for  $\mu_{\Delta,R}$. Examples at level 2 are given by (\ref{K2}) and (\ref{K11}), examples at level 4 are (examples up to level 6 can be found in the Appendix)
\be
\Big< K_{[3,1]}\cdot S_R \Big> =
\Big< \Big(P_3P_1-3P_2\underline{-3NP_1^2+3N^2}\Big)\cdot S_R \Big> = \mu_{[3,1],R}\cdot \Big<S_R\Big>
\nn\\
\Big< K_{[2,1,1]}\cdot S_R \Big> =
\Big< \Big(P_2P_1^2-NP_2-(N^2+4)P_1^2  +N^3+2N\Big)\cdot S_R \Big> = \mu_{[2,1,1],R}\cdot \Big<S_R\Big>
\nn\\
\Big< K_{[1,1,1,1]}\cdot S_R \Big>=
\Big< \Big(P_1^4-6NP_1^2  +3N^2\Big)\cdot S_R \Big> = \mu_{[1,1,1,1],R}\cdot \Big<S_R\Big>
\nn\\
\Big<K_{[4]}\cdot S_R\Big> =
\Big<(P_4-4NP_2-2P_1^2+2N^3+N)\cdot S_R\Big> = \mu_{[4],R}\cdot \Big< S_R\Big>
\nn \\
\Big<K_{[2,2]}\cdot S_R\Big>=
\Big<\left(P_2^2-2(N^2+2)P_2 +N^2(N^2+2)\right)\cdot S_R\Big>
= \mu_{[2,2],R}\cdot \Big< S_R\Big>
\label{complemid}
\ee
The underlined term could be eliminated with the help of (\ref{K11}),
but this causes an $N$-dependent shift of the eigenvalue
$\mu_{[3,1]}\longrightarrow \mu_{[3,1]}-3N\mu_{[1,1]}$.
In (\ref{complemid}) {\it per se} all $\mu_R$ are independent of $N$.
However, one can use (\ref{K2}) instead in order to remove the second and forth terms in this formula:
this would give rise to the $N$-independent shift $\mu_{[3,1]}\longrightarrow \mu_{[3,1]}+3\mu_{[2]}$.

Note that two equations of (\ref{WonS1}) can be compared with the corresponding equations from
this list, they differ by adding lower $N$-independent averages so that $\lambda_{[2]}=\mu_{[4]}+\mu_{[1,1]}$ and $\lambda_{[1,1]}=
\mu_{[2,2]}+\mu_{[2]}$. Generally, the identification for the part of relations that can be generated by the
$W$-operators is
$\lambda_{\Delta,R} = \mu_{2\Delta,R}+lower\ terms$,
where $2\Delta$ denotes a Young diagram with
all line lengths doubled, $2\Delta:=\{2\Delta_1\geq 2\Delta_2 \geq \ldots \ldots 2\Delta_{l_\Delta}\}$.

\section{Values of $\mu_{\Delta,R}$}

Finding the ``eigenvalues'' $\mu_{\Delta,R}$ is a separate challenge. As we explain in the next section, there is a general formula for them. However, the formula is not that simple, and it is instructive to look at examples.
The list of the first few is in the Table.

\bigskip

{\footnotesize
\hspace{-1.5cm}\begin{tabular}{|c|cc|ccccc|ccccccccccc|}
\hline
&&&&&&&&&&&&&&&&&&\\
R$\Big\backslash\Delta$& {[2]}& ${[1^2]}$& {[4]} &{[3,1]}&${[2^2]}$&${[2,1^2]}$&${[1^4]}$&{[6]}&{[5,1]}&{[4,2]}
&${[4,1^2]}$&${[3^2]}$&{[3,2,1]}&${[2^3]}$&${[3,1^3]}$&${[2^2,1^2]}$&${[2,1^4]}$&${[1^6]}$ \\
&&&&&&&&&&&&&&&&&&\\
\hline
${[2]}$        &{\bf 2}& 2  &   & & &  &   & &&&&& &&&&&\\
${[1^2]}$      &{\bf 2}& -2 &   &  &&  &  &  &&&&&&&&&&\\
\hline
${[4]}$         &{\bf 4}&4   & {\bf 8} & 8 & 8 & 8 & 8 &&&&&&&&&&&\\
${[3,1]}$         &{\bf 4}&-4   & {\bf 8} & 0 & 8 & -8 & -24 &&&&&&&&&&&\\
${[2^2]}$         &4&0   & {\bf 0} & -4 & 8 & 0 & 8 &&&&&&&&&&&\\
${[2,1^2]}$        &4&4  & -8 & 0 & 8 & 8 & -24 &&&&&&&&&&&\\
${[1^4]}$        &4&-4 & -8  & 8 & 8 & -8 & 8 &&&&&&&&&&&\\
\hline
${[6]}$                 &{\bf 6}&6   & {\bf 24} & 24 & 24 & 24 & 24 & {\bf 48}&48&48&48&48&48&48&48&48&48&48\\
${[5,1]}$               &{\bf 6}&-6   & {\bf 24} & 0 & 24 & -24 & -72 & {\bf 48}&0&48&-48&48&0&48&-96&-48&-144&-240\\
${[4,2]}$               & 6&2   & {\bf 8} & 0 & 24 & 8 & 24 & {\bf 0}&-16&16&-16&0&0&48&0&16&48&144\\
${[4,1^2]}$             &6&6   & 0 & 12 & 24 & 24 & -24 & -24&0&0&0&-24&24&48&-24&48&-48&-240\\
${[3^2]}$               &6&-2   & 8 & -8 & 24 & -8 & -8 & 0&0&16&16&-32&-16&48&16&-16&-16&-80\\
${[3,2,1]}$              &- & -   & - & - & - & - & - & -&-&-&-&-&-&-&-&-&-&-\\
${[2^3]}$             &6&2   & -8 & -8 & 24 & 8 & -8 &0 &0&-16&16&32&-16&48&-16&16&-16&80\\
${[3,1^3]}$           &6&-6   & 0 & 12 & 24 & -24 & -24 & -24&0&0&0&24&24&48&24&-48&-48&240\\
${[2^2,1^2]}$           &6&-2   & -8 & 0 & 24 & -8 & 24 & 0&16&-16&-16&0&0&48&0&-16&48&-144\\
${[2,1^4]}$         &6&6   & -24 & 0 & 24 & 24 & -72 & 48 &0&-48&-48&-48&0&48&96&48&-144&240\\
${[1^6]}$       &6&- 6   & -24 & 24 & 24 & -24 & 24 & 48&-48&-48&48&-48&48&48&-48&-48&48&-48\\
\hline
\end{tabular}
}

\bigskip

Clearly, transposition of $R$ preserves the absolute value of $\mu$:
\be
\mu_{\Delta,R^\vee} = (-1)^{l_\Delta+|\Delta|/2} \mu_{\Delta,R}
\label{transp}
\ee
We will prove it in the next section.

 In fact, the quantities in the Table are given by product formulas, in particular:
 \be
 \mu_{[2m],[2r]} =
\frac{\Big<K_{[2m]}\, S_{[2r]}\Big>}{\Big<S_{[2r]}\Big>} =\frac{S_{[2r-2m]}}{S_{[2r]}}\{\delta_{k,2}\}
 = \frac{(2r)!!}{(2r-2m)!!} \nn \\
 \mu_{[2m],[2r-1,1]} = \frac{\Big< K_{[2m]}\, S_{[2r-1,1]}\Big>}{\Big<S_{[2r-1,1]}\Big>}
 =\frac{S_{[2r-2m-1,1]}}{S_{[2r-1,1]}}\{\delta_{k,2}\} = \frac{(2r)!!}{(2r-2m)!!}
&  {\rm  for}\  r>2
\nn \\
 \mu_{[2m],[2r-2,2]} = \frac{\Big< K_{[2m]}\, S_{[2r-2,2]}\Big>}{\Big<S_{[2r-2,2]}\Big>}
 =\frac{S_{[2r-2m-2,2]}}{S_{[2r-2,2]}}\{\delta_{k,2}\} = \frac{(2r-2)!!}{(2r-2m-2)!!}
&  {\rm  for}\  m\geq 2
\nn \\
 \ldots
\ee
These $\mu_{\Delta,R}$ are shown boldfaced in the table. Expression through the values of Schur functions at $p_k=\delta_{k,2}$ are explained in (\ref{ev}) below.

All averages $\left<K_Q \cdot S_{[3,2,1]}\right>=0$, because $\left<S_{[3,2,1]}\right>=0$,
this is, in turn, because the factor $S_{[3,2,1]}\{\delta_{k,2}\} = 0$  in (\ref{siherm}).
Therefore, the corresponding $\mu_{Q,[3,2,1]}$ are not defined.
The same is true for all $S_{[\ldots4321]}$, which are independent of even time-variables.
In fact, it is sufficient for vanishing of the average that the Schur polynomials
does not contain the item $p_2^{|R|/2}$, this happens for $S_{[5,2,1]}$, $S_{[1,1,1,2,3]}$
and a number of other examples of bigger sizes.

The table has clearly a triangle structure, since, if $|\Delta|>|R|$, the corresponding $W^-_\Delta$ contains more derivatives than the degree of $H$ in $S_R$.

\bigskip

The first example is provided by $\mu_{[1,1],R}$ at level $2$.
While
\be
\mu_{[2],R} =\lambda_{[1],R} = |R| \nn
\ee
is very simple, expression for $\mu_{[1,1],R}$ is quite involved:
it depends on the number $l_R$ of columns in the diagram
$R=\left[r1\geq r_2\geq r_3\geq\ldots \geq r_{l_R}>0\right]$:
\be
\mu_{[1,1],[r_1]}&=&r_1\cdot P_e,\nn\\
\mu_{[1,1],[r_1,r_2]}&=& (r_1-r_2) \cdot P_{ee} + (-r_1+r_2-2)\cdot P_{oo}\nn\\
\mu_{[1,1],[r_1,r_2,r_3]}&=&(r_1-r_2+r_3)\cdot P_{eee} +(-r_1+r_2+r_3-2)\cdot P_{ooe}  + (r_1+r_2-r_3+2)\cdot P_{eoo}\nn\\
\mu_{[1,1],[r_1,r_2,r_3,r_4]}&=&(r_1-r_2+r_3-r_4) \cdot P_{eeee}
+ (-r_1+r+2+r+3-r_4-2)\cdot P_{ooee}+\nn\\ &+& (r_1+r_2-r_3-r_4+2)\cdot P_{eooe}
+ (r_1-r_2-r_3+r_4-2) \cdot P_{eeoo}+\nn\\   &+& (-r_1-r_2+r_3+r_4-6) \cdot P_{oeeo} + (-r_1+r_2-r_3+r_4-4) \cdot P_{oooo}
\ee
Projector $P_{eoo}$ here, for instance, means that $r_1$ is even,  $r_2$ and $r_3$ are odd so that
all the values of $|R|$ are even.
Averages of the type $oeo$, $eoeo$ and $oeoe$ are all vanishing.

Different lines in these formulas are connected smoothly,
one should just pick up the terms with $E$ at the very right position and
put the highest $r_{l_R}=0$.

\section{Derivation of (\ref{W}) and explicit formula for $\mu_{\Delta,R}$}

Alternative representation of $\mu_{[1,1],R}$ can be deduced from (\ref{yesavef}), which we are going to derive now.
The first examples of this relation are
\be
\left(\tr \frac{\p}{\p H}\right)^2 S_{[2]} =
\left(\tr \frac{\p}{\p H}\right)^2 \left(\frac{\tr H^2+(\tr H)^2}{2}\right)=N(N+1)
\nn \\
\left(\tr \frac{\p}{\p H}\right)^2 S_{[1,1]} =
\left(\tr \frac{\p}{\p H}\right)^2 \left(\frac{-\tr H^2+(\tr H)^2}{2}\right)
=N(N-1)
\nn \\
\left(\tr \frac{\p}{\p H}\right)^2 S_{[k]}=(N+k-2)(N+k-1)S_{[k-2]}
\nn \\
\left(\tr \frac{\p}{\p H}\right)^2 S_{[k_1,k_2]}=(N+k_1-1)(N+k_1-2)S_{[k_1-2,k_2]}
+2(N+k_1-1)(N+k_2-2)S_{[k_1-1,k_2-1]}
+ \nn\\
+(N+k_2-1)(N+k_2-2)S_{[k_1,k_2-2]}
\nn \\
\ldots
\ee

\noindent
Clearly, these formulas are obtained by successive application of the operator
\be
\left(\tr \frac{\p}{\p H}\right) S_{R}=\sum_{\Box}(N+j-i)S_{R-\Box_{i,j}}
\label{ddH}
\ee
where the box $\Box_{i,j}$ with coordinates $(i,j)$ is removed from the Young diagram $R$ so that $R-\Box_{i,j}$ is still a Young diagram. This formula is a kind of inverse of the Pieri rule.

{\bf Despite this time the operator changes $S_R$}, which is no longer its eigenfunction,
like it was in (\ref{SefW}),
{\bf it does not change the average!}
This is because of the very special coefficient at the r.h.s. of (\ref{ddH}).
Indeed, note that
\be\label{prod}
{S_R\{N\}\over d_R}=\prod_{\Box_{i,j}\in R}(N+j-i)
\ee
and hence
\be
(N+j-i) = {S_R\{N\}\over S_{R-\Box_{i,j}}\{N\}}\cdot \frac{d_{R-\Box_{i,j}}}{d_R}
\ee
This means that
\be
\Big< (P_1^2-N)\cdot S_R \Big>
\stackrel{(\ref{complemid})}{=}
\left<\left(\tr \frac{\p}{\p H}\right)^2 S_R\right>
= \nn \\
\stackrel{(\ref{ddH})}{=}
\sum_{\Box_1,\Box_2\in R}
{S_R(N)\over S_{R-\Box_1-\Box_2}(N)}\cdot \frac{d_{R-\Box_1-\Box_2}}{d_R}
\cdot\Big<S_{R-\Box_1-\Box_2} \Big>
\ \stackrel{(\ref{siherm})}{=}
\mu_{[1,1],R}\cdot\Big<S_R\Big>
\ee
with
\be
\mu_{[1,1],R}={1\over S_R\{\delta_{k,2}\}}\sum_{\Box_1,\Box_2\in R} S_{R-\Box_1-\Box_2}\{\delta_{k,2}\}
\ee
The sum over the boxes of the Young diagram is such that the diagram obtained after removing
any of these two boxes and both of them still remains a Young diagram.
When there are two different ways to achieve the final state, a combinatorial coefficient $2$ appears.
For example,
\be
\mu_{[1,1],[7,1]} = \frac{S_{[5,1]}+2S_{[6]}}{S_{[7,1]}}\{\delta_{k,2}\}
\nn \\
\mu_{[1,1],[8,4]} = \frac{S_{[6,4]}+2S_{[7,3]}+S_{[8,2]}}{S_{[8,4]}}\{\delta_{k,2}\}
\nn \\
\ldots
\ee
When $S_R\{\delta_{k,2}\}=0$, the corresponding Gaussian average vanishes and $\mu_{[1,1],R}$ is not defined.
One can check that in these cases the numerator vanishes as well.

One can similarly consider the action higher (even) degrees of operator $\left(\tr \frac{\p}{\p H}\right)^{2n}$, in this case with the same line of reasoning, one obtains
\be
\mu_{[1^{2n}],R}={1\over S_R\{\delta_{k,2}\}}\sum_{\{\Box_i\}\in R} S_{R-\sum_i^{2n}\Box_i}\{\delta_{k,2}\}
\ee
where one has to remove $2n$ boxes from the Young diagram preserving the property of being the Young diagram.

Furthermore,
\ytableausetup {boxsize=0.4em}
\be
W^-_{2}S_R=\tr\left( \frac{\p^2}{\p H^2}\right) S_{R}=\sum_{\ydiagram{2}}c_{_{\ydiagram{2}}}S_{R-\ydiagram{2}}-
\sum_{\ydiagram{1,1}}c_{_{\ydiagram{1,1}}}S_{R-\ydiagram{1,1}}
\label{d2H}
\ee
where the two boxes are removed from rightmost part of the $R$ in such a way that obtained is still a Young diagram.
The general formula is

\be\label{iP}
W^-_{k}S_R=\tr\left( \frac{\p^k}{\p H^k}\right) S_{R}=\sum_{h_k}(-1)^{{\rm ht}(h_k)}c_{_{R,h_k}}S_{R-h_k}
\ee
where $h_k$ is a border strip of length $k$ (i.e. containing $k$ boxes) \cite{Mac}, ${\rm ht}(h_k)$ is its height (defined to be one less than the number of rows it occupies), the sum runs over all such border strips, and all the coefficients $c_{_{R,h_k}}$ are products of $N-i+j$ over subtracted boxes. Note that the border strips are called rim hooks in \cite{Pop}. In order to get $W^-_\Delta$ with a few rows, one has to apply (\ref{iP}) sequentially.

Formula (\ref{iP}) is a kind of inverse to the Pieri rule
\be
p_kS_R=\sum_{Q:\ Q-R=h_k}(-1)^{{\rm ht}(h_k)}S_Q
\ee 

Now one immediately obtains the explicit formula for $\mu_{\Delta,R}$:
\be\label{ev}
\boxed{
\mu_{\Delta,R}={1\over S_R\{\delta_{k,2}\}}\sum_{\{h_{\Delta_a}\}\in R}
(-1)^{{\rm ht}(h_{\Delta_a})}S_{R-\sum_ah_{\Delta_a}}\{\delta_{k,2}\}
}
\ee
where $h_{\Delta_a}$ are the border strips removed from the Young diagram $R$ in accordance with (\ref{iP}), with the corresponding sign taken into account.

Note that, since $S_{R^\vee}\{p_k\}=(-1)^{|R|}S_R\{-p_k\}$ and $S_R\{p_k=\delta_{k,2}\}\sim p_2^{|R|/2}$, one immediately obtains from this formula relation (\ref{transp}). Indeed, let us look, for instance, at the case of $K_{[2]}$. Then,
\be\label{mu2}
\mu_{[2],R}={1\over S_R\{\delta_{k,2}\}}\left(\sum_{\ydiagram{2}\in R}S_{R-\ydiagram{2}}\{\delta_{k,2}\}-
\sum_{\ydiagram{1,1}\in R}S_{R-\ydiagram{1,1}}\{\delta_{k,2}\}\right)
\ee
and every term in the sum gets the sign $(-1)^{|\Delta|/2}$ under conjugation of the Young diagram. Besides, in the course of this conjugation, one has to permute the two terms in (\ref{mu2}). This gives total sign $(-1)^{|\Delta|/2+1}$. This case is certainly very trivial, because the difference of two sums (\ref{mu2}) is equal to $|R|$.

For the generic $K_\Delta$, there is a sign factor $(-1)^{|\Delta|}$ from the ratios of $S_{R-\Delta}/S_R$ and an additional factor of $(-1)^{l_\Delta}$ that follows from the fact that the conjugation of $R$ changes additionally the sign of $\mu_{\Delta,R}$ for every even part of partition $\Delta$ because of formula (\ref{iP}) (the height of the border strip changes the parity under conjugation in this case). This gives $(-1)^{|\Delta|+\#_e}$, where $\#_e$ is the number of even parts of partition $\Delta$. However, as soon as the number of odd parts (which do not give rise to this additional sign changing) is even, this formula can be changed for $(-1)^{|\Delta|+l_\Delta}$.

\section{Conclusion}

To conclude, we have found a new indirect implication of superintegrability:
factorization of peculiar pair correlators,
where one component of the pair is just the character,
while the other one is its $N$-dependent deformation. This factorization is due to existence
of an infinite set of commuting operators $W^-_\Delta$, (\ref{Wd}) that provide a mapping from the space ${\cal X}$ of 
invariant matrix polynomials to the space ${\cal Z}$ of polynomials of matrix size, $F_{W_\Delta^-}:\ {\cal X}\longrightarrow {\cal Z}$. This mapping is manifestly given by the matrix averaging
\be
\forall f\in {\cal X}:\  F_{W_\Delta^-}(f)=\Big<W_\Delta^-\cdot f\Big>\in {\cal Z}
\ee 
The space ${\cal Z}$ can be spanned (ambiguously) by matrix averages of the Schur functions $\Big<S_R\Big>$. As we demonstrated in this paper, the Schur functions turn out to be eigenvectors of the mapping:
\be\label{Wm}
F_{W_\Delta^-}(S_R)=\Big<W_\Delta^-\cdot S_R\Big>=\mu_{\Delta,R}\Big<S_R\Big>
\ee
while, on the space ${\cal X}$, i.e. before the averaging, this is not the case, since the action of $W_\Delta^-$ on $S_R$ decrease its grading to $|R|-|\Delta|$.

It is an open question, if this construction persists in other
superintegrable models, which could help to understand if it is in fact a {\it direct} corollary of SI or not.
Note also that, in other models, the space ${\cal Z}$ may need an extension to the space of rational functions of the matrix size, see, for instance, the case of two-logarithm (Selberg) models, \cite{MMrev}.

Also an important issue is what are interrelations between the set of operators $W_\Delta^-$ and a commutative set of generalized cut-and-join operators $W_\Delta$. Indeed, the Schur function are the eigenvectors of the latter already on the space ${\cal X}$.
However, as a price of it, if one rewrites (\ref{Wm}) as a statement about the pair correlator (\ref{W}),
\be
\Big<K_\Delta\cdot S_R\Big>=\mu_{\Delta,R}\cdot  \Big<S_R\Big>
\ee
the set of $W_\Delta$ generates too little set of polynomials $K_\Delta$, while the set of $W_\Delta^-$ gives rise to the complete basis of $K_\Delta$. The operators $W_\Delta^-$ are of a kind of lowering operators in an algebra, and $W_\Delta$ are Cartan-like operators in it. Indeed, as one can notice, the operators $W_k^-$ looks as positive harmonics in the $w_\infty$-algebra, while $W_k$, as zeroth harmonics in it: as $\hat W^{(k+1)}_k$ and $\hat W^{(k+1)}_0$ correspondingly in terms of \cite[secs.8-9]{AMMN}.  Note that these zeroth harmonic operators can generate the Virasoro and $W$-algebra constraints \cite{MMMR}.

\section*{Acknowledgements}

Our work is partly supported by the grant of the Foundation for the Advancement of Theoretical Physics ``BASIS'', by joint grants 21-52-52004-MNT-a and 21-51-46010-ST-a.

\section*{Appendix}

We list here the first few operators  $K_\Delta$. They celebrate the property that their bilinear averages with all $S_R$, are proportional to the averages of $S_R$:
$\Big<K_\Delta S_R\Big> = \mu_{\Delta,R}\Big<S_R\Big>$. However, this condition does not fix $K_\Delta$ unambiguously, therefore we use a precise definition:  $K_\Delta$ are obtained by the action of operators $\hat W^-\Delta$  on the Gaussian weight $e^{-\frac{1}{2}\tr H^2}$, see eq.(\ref{KD}). Then,

\be
K_{[2]}=P_2-N^2\nn\\
K_{[1,1]}=P_1^2-N
\ee
\be
K_{[4]}=P_4-4NP_2-2P_1^2+N(2N^2+1)
\nn \\
 K_{[3,1]}=P_3P_1-3P_2-3NP_1^2+3N^2
\nn\\
K_{[2,2]}=P_2^2-2(N^2+2)P_2 +N^2(N^2+2)\nn\\
 K_{[2,1,1]}=P_2P_1^2-NP_2-(N^2+4)P_1^2  +N(N^2+2)
\nn\\
 K_{[1,1,1,1]}=P_1^4-6NP_1^2  +3N^2
\ee
\be
K_{[6]}=
P_6-6NP_4-6P_1P_3-3P_2^2+15(N^2+1)P_2+15NP_1^2-5N^2(N^2+2)\nn\\
K_{[5,1]}=
P_5P_1-5P_4-5NP_1P_3-3P_2^2-5P_1^2P_2+20NP_2+5(2N^2+3)P_1^2-5N(2N^2+1)\nn\\
K_{[4,2]}=
P_4P_2-(N^2+8)P_4-4NP_2^2-2P_1^2P_2+(6N^3+25N)P_2+2(N^2+6)P_1^2-N(2N^2+1)(N^2+4)\nn\\
K_{[4,1,1]}=
P_4P_1^2-NP_4-8P_1P_3-4NP_1^2P_2-2P_1^4+4(N^2+3)P_2+(2N^3+27N)P_1^2-N^2(2N^2+13)\nn\\
K_{[3,3]}=
P_3^2-9P_4-6NP_1P_3+27NP_2+9(N^2+1)P_1^2-3N(4N^2+1)\nn\\
K_{[3,2,1]}=
P_3P_2P_1-(N^2+8)P_1P_3-3P_2^2-3NP_1^2P_2+6(N^2+3)P_2+3N(N^2+6)P_1^2-3N^2(N^2+4)\nn\\
K_{[2,2,2]}=
P_2^3-3(N^2+4)P_2^2+3(N^2+2)(N^2+4)P_2-N^2(N^2+2)(N^2+4)\nn\\
K_{[3,1,1,1]}=
P_3P_1^3-3NP_3P_1-9P_2P_1^2-3NP_1^4+9NP_2+18(N^2+1)P_1^2-3N(3N^2+2)\nn\\
K_{[2,2,1,1]}=
P_2^2P_1^2-NP_2^2-2(N^2+6)P_2P_1^2+2N(N^2+4)P_2+(N^2+4)(N^2+6)P_1^2-N(N^2+2)(N^2+4)\nn\\
K_{[2,1,1,1,1]}=
P_2P_1^4-6NP_2P_1^2-(N^2+8)P_1^4+3N^2P_2+6N(N^2+6)P_1^2-3N^2(N^2+4)\nn\\
K_{[1,1,1,1,1,1]}=
P_1^6-15NP_1^4+45N^2P_1^2-15N^3
\nn\\
\ee


\begin{thebibliography}{12}

\bibitem{IMM} H.~Itoyama, A.~Mironov, A.~Morozov,
  JHEP {\bf 1706} (2017) 115,
arXiv:1704.08648

\bibitem{MM} A.~Mironov, A.~Morozov,
  Phys.\ Lett.\ {\bf B771} (2017) 503,
arXiv:1705.00976

\bibitem{DiF} P.~Di Francesco, C.~Itzykson, J.~B.~Zuber,
  Commun.\ Math.\ Phys.\  {\bf 151} (1993) 193,
  hep-th/9206090

\bibitem{IdF} P.~Di Francesco, C.~Itzykson,
Ann. Inst. H. Poincare Phys. Theor. \textbf{59} (1993) 117-140,
hep-th/9212108

\bibitem{Ivan1} I.~K.~Kostov, M.~Staudacher,
Phys. Lett. \textbf{B394} (1997) 75-81,
hep-th/9611011

\bibitem{Ivan2} I.~K.~Kostov, M.~Staudacher, T.~Wynter,
Commun. Math. Phys. \textbf{191} (1998) 283-298,
hep-th/9703189

\bibitem{Orlov} A. Orlov, 
Int. J. Mod. Phys. {\bf A19, supp02} (2004) 276-293, nlin/0209063

\bibitem{MMSh} A.~Mironov, A.~Morozov, S.~Shakirov,
JHEP \textbf{02} (2011) 067,
arXiv:1012.3137

\bibitem{MMShS} A.~Mironov, A.~Morozov, S.~Shakirov, A.~Smirnov,
Nucl. Phys. \textbf{B855} (2012) 128-151,
arXiv:1105.0948

\bibitem{AMMN} A. Alexandrov, A. Mironov, A. Morozov, S. Natanzon,
JHEP 11 (2014) 080,  arXiv:1405.1395

\bibitem{Orlov2}  S. Natanzon, A. Orlov, arXiv:1407.8323

\bibitem{Pop} C.~Cordova, B.~Heidenreich, A.~Popolitov, S.~Shakirov,
  Commun.\ Math.\ Phys.\  {\bf 361} (2018)   1235,
  arXiv:1611.03142

\bibitem{MMten} A.~Mironov, A.~Morozov,
  Phys.\ Lett.\ {\bf B774} (2017) 210,
arXiv:1706.03667

\bibitem{MPS} A.~Morozov, A.~Popolitov and S.~Shakirov,
  Phys.\ Lett. {\bf B784} (2018) 342,
  arXiv:1803.11401

    \bibitem{MMsum} A.~Mironov, A.~Morozov,
  JHEP {\bf 1808} (2018) 163,
arXiv:1807.02409

\bibitem{MMell}   A.~Mironov, A.~Morozov,
Phys. Lett. \textbf{B816} (2021), 136196,
arXiv:2011.01762; ibid.,
136221,
arXiv:2011.02855

\bibitem{MMkon} A.~Mironov, A.~Morozov,
Eur. Phys. J. \textbf{C81} (2021) 270,
arXiv:2011.12917

\bibitem{MMl} A.Mironov, A.Morozov,  Phys.Lett. {\bf B816} (2021) 136268,  arXiv:2102.01473

\bibitem{Zabz} L.~Cassia, R.~Lodin, M.~Zabzine,
JHEP \textbf{10} (2020) 126,
arXiv:2007.10354

\bibitem{MMPstud} A.~Mironov, A.~Morozov, A.~Popolitov,
Phys. Lett. \textbf{B824} (2022) 136833,
arXiv:2107.13381

\bibitem{WWWZ} L.~Y.~Wang, R.~Wang, K.~Wu, W.~Z.~Zhao,
Nucl. Phys. \textbf{B973} (2021) 115612,
arXiv:2110.14269

\bibitem{MMMZh} A.~Mironov, V.~Mishnyakov, A.~Morozov, A.~Zhabin,
arXiv:2112.11371

\bibitem{MMrev} A.~Mironov, A.~Morozov,
arXiv:2201.12917

\bibitem{MMZ} A.~Mironov, A.~Morozov, Z.~Zakirova,
Phys. Lett.  \textbf{B831} (2022) 137178,
arXiv:2203.03869

\bibitem{MO} V. Mishnyakov, N. Oreshina,  arXiv:2203.15675

\bibitem{Kaz} V.A. Kazakov, M. Staudacher, T. Wynter, 
hep-th/9601153, 1995 Carg\`ese Proceedings

\bibitem{Ramg} S. Corley, A. Jevicki, S. Ramgoolam, Adv.Theor.Math.Phys. {\bf 5} (2002) 809-839, hep-th/0111222

\bibitem{KPSS} C. Kristjansen, J. Plefka, G. W. Semenoff, M. Staudacher, Nucl.Phys. {\bf B643} (2002) 3-30, hep-th/0205033

\bibitem{BEM} M. Tierz, Mod. Phys. Lett. A19 (2004) 1365-1378, hep-th/0212128\\
A. Brini, B. Eynard, M. Mari\~no, Annales Henri Poincar\'e. Vol. 13. No. 8. SP Birkh\''{a}user Verlag Basel, 2012, arXiv:1105.2012

\bibitem{MKR} R. de Mello Koch, S. Ramgoolam, arXiv:1002.1634

\bibitem{MMd} A. Mironov, A. Morozov, {\it AGT and Nekrsov calculus as a corollary of superintegrability}, to appear

\bibitem{Kadell} K.W.J. Kadell,
Compositio Math. {\bf 87} (1993) 5-43

\bibitem{MMN} A.~Mironov, A.~Morozov, S.~Natanzon,
Theor. Math. Phys. \textbf{166} (2011) 1-22,
arXiv:0904.4227;
J. Geom. Phys. \textbf{62} (2012) 148-155,
arXiv:1012.0433

\bibitem{Mac} I.G. Macdonald,
{\it Symmetric functions and Hall polynomials}, Second Edition, Oxford University Press,
1995

\bibitem{MMMR} A.~Mironov, V.~Mishnyakov, A.~Morozov, R.~Rashkov,
Eur. Phys. J. \textbf{C81} (2021) 1140,
arXiv:2105.09920


\end{thebibliography}
\end{document}